\documentclass{article}

\usepackage{arxiv}

\usepackage[utf8]{inputenc} 
\usepackage[T1]{fontenc}    
\usepackage{hyperref}       
\usepackage{url}            
\usepackage{booktabs}       
\usepackage{amsfonts}       
\usepackage{nicefrac}       
\usepackage{microtype}      
\usepackage{lipsum}		
\usepackage{graphicx}
\usepackage{doi}

\usepackage{graphicx}
\usepackage{url}
\usepackage{mdframed}
\usepackage{listings}
\usepackage{caption}
\usepackage{amsmath,amsfonts,amssymb}
\usepackage[ruled,vlined]{algorithm2e}

\usepackage{microtype}
\usepackage[T1]{fontenc}

\definecolor{deepblue}{rgb}{0,0,0.5}
\definecolor{deepred}{rgb}{0.6,0,0}
\definecolor{deepgreen}{rgb}{0,0.5,0}
\definecolor{codegreen}{rgb}{0,0.6,0}
\definecolor{codegray}{rgb}{0.5,0.5,0.5}
\definecolor{codepurple}{rgb}{0.58,0,0.82}
\definecolor{backcolour}{rgb}{0.95,0.95,0.92}

\newcommand\pythonstyle{\lstset{
    language=Python,
    basewidth=0.5em,
    basicstyle=\ttfamily\footnotesize,
    morekeywords={self,as},     
    keywordstyle=\color{deepgreen},
    keywordstyle={[2]\color{blue}},
    keywords=[2]{incorporate_data,initialize,parallel_svd,parallel_qr,linalg.svd,linalg.qr,matmul,concatenate,range,append,axis,diag},
    stringstyle=\color{deepred},
    commentstyle=\color{codegray},
    numberstyle=\tiny\color{codegray},
    escapechar={|},
    basicstyle=\ttfamily\footnotesize,
    frame=tb,       
    showstringspaces=true, 
    belowcaptionskip=0.2cm
}}

\lstnewenvironment{python}[1][]
{
\pythonstyle
\lstset{#1}
}
{}

\usepackage{graphicx}
\usepackage{subfigure}

\title{PyParSVD: A streaming, distributed and randomized singular-value-decomposition library}

\author{Romit Maulik \\
	Mathematics and Computer Science Division\\
	Argonne National Laboratory\\
	Lemont, IL-60439, USA \\
	\texttt{rmaulik@anl.gov} \\
	\And
	Gianmarco Mengaldo \\
	Department of Mechanical Engineering\\
	National University of Singapore\\
	Singapore \\
	\texttt{mpegim@nus.edu.sg} \\
}

\renewcommand{\shorttitle}{PyParSVD: A streaming, distributed and randomized singular-value-decomposition library}

\begin{document}
\maketitle

\begin{abstract}
	We introduce PyParSVD\footnote{https://github.com/Romit-Maulik/PyParSVD}, a Python library that implements a streaming, distributed and randomized algorithm for the singular value decomposition. To demonstrate its effectiveness, we extract coherent structures from scientific data. Futhermore, we show weak scaling assessments on up to 256 nodes of the Theta machine at Argonne Leadership Computing Facility, demonstrating potential for large-scale data analyses of practical data sets.
\end{abstract}

\section{Introduction}\label{S:1}

The singular value decomposition (SVD) is a matrix factorization technique that decomposes a general complex $M \times N$ matrix $\mathbf{A} \in \mathbb{C}^{M\times N}$ into three matrices, $\mathbf{U} \in \mathbb{C}^{M\times M}$, $\boldsymbol{\Sigma} \in \mathbb{C}^{M\times N}$, and $\mathbf{V} \in \mathbb{C}^{N\times N}$, that is 
\begin{equation}
    \mathbf{A} = \mathbf{U}\boldsymbol{\Sigma}\mathbf{V}^{*}
\end{equation}
where the symbol $^{*}$ denotes a matrix's complex conjugate transpose, $\mathbf{U}$ and $\mathbf{V}$ are complex unitary matrices, $\boldsymbol{\Sigma}$ is a rectangular diagonal matrix, and $\mathbb{C}$ denotes the space of complex numbers. The diagonal entries of $\boldsymbol{\Sigma}$ are known as the singular values of $\mathbf{A}$, while the columns of $\mathbf{U}$ and $\mathbf{V}$ are known as the left and right singular vectors of $\mathbf{A}$. For many problems, the SVD is applied to real matrices, that is $\mathbf{A} \in \mathbb{R}^{M\times N}$, thereby yielding $\mathbf{A} = \mathbf{U}\boldsymbol{\Sigma}\mathbf{V}^{\text{T}}$, where $^{\text{T}}$ represents a matrix's transpose, and $\mathbb{R}$ is the space of real numbers. 

The SVD was introduced with a series of original contributions by Eugenio Beltrami, Camille Jordan, James Joseph Sylvester, Ernhard Schmidt, and Hermann Weyl in the 19$^{th}$ and early 20$^{th}$ century -- the interested reader can refer to  Stewart~\cite{stewart1993early} for the early history of SVD -- and it is today at the hearth of virtually any data analysis and matrix computation framework. It can be used to identifying patterns in high-dimensional datasets, for solving least-square problems, and calculating a matrix pseudoinverse, to cite just a few. It is therefore not surprising that several algorithms to compute the SVD efficiently have been presented in the literature, from the pioneering contribution of Golub et al.,~\cite{golub1965calculating,golub1971singular}, to its generalizations and improvements~\cite{van1976generalizing,klema1980singular,chan1982improved,gu1993stable,jessup1994parallel,gu1995downdating}. For a review of existing algorithms to compute the SVD and their pros and cons, the interested reader is referred to Menon and Elkan~\cite{menon2011fast}, and to Dongarra et al.~\cite{dongarra2018singular}. More recently, extensions of the SVD to multidimensional tensors have been proposed~\cite{de2000multilinear,grasedyck2010hierarchical}.

The SVD has gained its tremendous popularity thanks to the relative simplicity, and to its connections to $n$-dimensional geometry. The latter allows for a visual interpretation of the SVD decomposition, that is particularly useful when making sense of complex multidimensional data. However, the enormous amount of data that has become available in recent years have created the critical need for ever-than-before fast SVD algorithms. Within this drive are the efforts towards efficient parallel SVD~\cite{wang2016approximate}, randomized SVD~\cite{Erichson_2017_ICCV}, and quantum SVD~\cite{rebentrost2018quantum}. In this paper, we leverage streaming algorithms, novel approximate partitioned methods, and randomization, to construct an efficient
streaming, parallel, randomized algorithm for the SVD. The algorithm is conveniently implemented in Python and provides a novel framework that can be used for the analysis of large datasets. It can also be ported in existing codes, where fast SVD computations are required, thereby providing an acceleration of the SVD computation step. The code has been extensively tested in several machines and can additionally serve as a starting point for novel fast SVD implementations in different programming languages, that may be required by users.

The rest of the paper is organized as follows. Section~\ref{S:2} highlights the needs for efficient and fast SVD algorithms. Section~\ref{S:3} introduces the algorithm underlying PyParSVD. Section~\ref{S:4} outlines the implementation of PyParSVD, along with parallel scaling results on reanalysis weather data. Finally, section~\ref{S:5} draws the main conclusions and future work.

\section{Why do we need the SVD?}
\label{S:2}

In recent years, the SVD has become a central tool in many practical applications involving matrix computation and data analysis, due to its simplicity and interpretability. Examples include modal decomposition and reduced order modeling, data compression, recommender and facial recognition systems~\cite{turk1991face, sarwar2002incremental}, matrix inversion~\cite{golub1965calculating}, least square fitting, and numerical calculation of matrix properties. The common underlying factors shared by these applications are 
\begin{enumerate}
\item the need for low rank (or reduced) representation of a matrix,
\item the need for a set of (orthogonal) bases for the row and column spaces of a matrix, and 
\item the need for information about the rank of a matrix.
\end{enumerate}

In terms of data analysis for example, it is possible to reduce the dimension of a data matrix $\mathbf{A}$ of rank $N$, by noting that it is possible to approximate it with another matrix $\widetilde{\mathbf{A}}$ of lower rank $r \ll N$. This is particularly useful for data compression, in the context of e.g.\ compressive sensing. A low-rank representation of the matrix $\mathbf{A}$ is also adopted in the context of reduced order modeling. In this case, one seeks to reduce the number of degrees of freedom of the original system described by matrix $\mathbf{A}$, and construct a lower-dimensional manifold for the problem. This is typically achieved by constructing suitable orthogonal bases via SVD and eigendecomposition, that can optimally represent the information content of the system, in a significantly more compact way. This, in turn, can reveal hidden coherent patterns in high-dimensional data. The proper orthogonal decomposition (POD)~\cite{chatterjee2000introduction}, along with its variants, including the spectral proper orthogonal decomposition (SPOD)~\cite{towne2018spectral,mengaldo2021pyspod}, are SVD-based techniques that allow identifying a reduced order representation of the original system by truncating the number of singular values (and consequently limiting the dimension of the singular vectors). The reduced representation is typically achieved in an orthogonal coordinate systems constituted by the POD (or SPOD) modes, calculated as the eigendecomposition of the covariance data matrix (or the cross-spectral density matrix in the case of the SPOD). These techniques have been widely used in fluid dynamics~\cite{chatterjee2000introduction}, as well as in weather and climate research (under the name empirical orthogonal function~\cite{lorenz1956empirical} and spectral empirical orthogonal functions analyses~\cite{schmidt2019spectral}) for data analysis and reduced order modeling. Similarly, the principal component analysis (PCA)~\cite{pearson1901liii,moore1981principal}, also known as Karhunen–Loève transform (KLT) or Hotelling transform, seeks to perform a change of coordinates or basis of the underlying matrix data representation. PCA is commonly used in machine learning for feature selection and data engineering. The terms PCA, POD, KLT and Hotelling transformations are frequently used interchangeably in the literature, and in many cases are synonyms. SVD-based techniques are also at the foundation of complementary and more recently developed data-driven analysis methods, such as the dynamic mode decomposition~\cite{schmid2010dynamic}, data-driven Koopman operator theory~\cite{li2017extended}, and sparse system identification~\cite{brunton2016discovering}. 

The SVD also serves as a platform for matrix theory and computation, and it has been used for several applications. In terms of matrix approximation, it can be used to calculate the pseudoinverse of a matrix $\mathbf{A}$, that is $\mathbf{A}^{\dag} = \mathbf{U}^{*}\boldsymbol{\Sigma}^{\dag}\mathbf{V}$, where $^{\dag}$ denotes pseudoinverse. The pseudoinverse of $\boldsymbol{\Sigma}$ is simply obtained by taking the reciprocal of the non-zero values on the diagonal, leaving zeros in place and transposing the matrix (that is in general rectangular). The pseudoinverse and its calculation via the SVD is used for linear least squares, such as the ordinary, weighted, and generalized least squares~\cite{merriman1877list, aitken1936iv}. Linear least squares are the basis for data fitting, and find applications in virtually every quantitative field, from engineering and applied science to economics and social science. 

Finally, but not least important, SVD is used in the context of matrix computation. In fact, SVD can be used to provide an explicit representation of range, null space, and rank of a generic matrix $\mathbf{A}$, with obvious implications in linear algebra. 
 
While there are several applications that use SVD algorithms in an effective and scalable way (see for example Google, Facebook, Netflix and Amazon, that use large-scale SVD computations for their recommender and image recognition systems), there is a pressing need to produce novel SVD algorithms that are lightweight, and highly scalable. This is useful for online SVD computations and in situations where there are limited computational resources. The streaming, randomized and parallel SVD presented here helps filling this gap, as it provides an efficient platform for lightweight SVD. This can be used for applications, where there is the need to compute the SVD on the fly or online.

\section{A streaming, distributed and randomized SVD algorithm}
\label{S:3}
The streaming, distributed, and randomized algorithm underlying the PyParSVD library, is constituted of three building blocks, (i) a streaming SVD in-situ~\cite{levy1998sequential}, a parallel distribution strategy based on approximate partitioned method of snapshots~\cite{wang2016approximate}, and (iii) randomized linear algebra. These three building blocks are described in the next three subsections, and their implementation in PyParSVD can be found in section~\ref{S:4}.

\subsection{Streaming SVD}
The first building block of our framework relies on Levy and Lindenbaums' method for performing a streaming SVD \cite{levy1998sequential} in-situ. One of our target applications is to use this SVD for analyzing the presence of coherent structures in data. Usually, this analysis is performed by constructing a data matrix $\mathbf{A} \in \mathbb{R}^{M \times N}$. $N$ refers to the number of `snapshots' of data collected for the analysis and $M$ is the number of degrees of freedom in each snapshot. For the purpose of this analysis, $M >> N$ and the regular SVD gives us
\begin{align}
    \mathbf{A} = \mathbf{U} \boldsymbol{\Sigma} \mathbf{V}^{\text{T}}
\end{align}
where $\mathbf{U} \in \mathbb{R}^{M \times N}, \boldsymbol{\Sigma} \in \mathbb{R}^{N \times N}, \mathbf{V} \in \mathbb{R}^{N \times N}$. The classical SVD computation scales as $O(MN^2)$ and requires $O(MN)$ memory. This analysis becomes intractable for computational physics applications such as computational fluid dynamics or numerical weather prediction, where the degrees of freedom may grow very large. In their seminal paper, Levy and Lindenbaum proposed a streaming variant of the SVD that reduces the computational and memory complexity significantly. It performs this by extracting solely the first $K$ left singular vectors, which correspond to the $K$ largest coherent structures. Consequently, we are able to reduce the cost of the SVD to $O(MNK)$ operations and the memory footprint also reduces to $O(MK)$. The streaming component of this technique consists of updating the left singular eigenvectors in a batch-like manner. We summarize the procedure in Algorithm \ref{Algo1}. The scalar forget factor \texttt{ff}, set between 0 and 1, controls the effect of older data batches on the final result for $\mathbf{U}_{i}$. Setting this value to 1.0 implies that the online-SVD converges to the regular SVD utilizing all the snapshots in one-shot. Setting values of \texttt{ff} less than one reduces the impact of the snapshots observed in previous batches of the past. We utilize an \texttt{ff}$=0.95$ for this study. A Python and NumPy implementation for this algorithm is shown in Listing \ref{LST1}.
\begin{algorithm}
\SetAlgoLined
\KwResult{Truncated left singular vector $\mathbf{U}_i$ after $i$ batch iterations.}
 \textbf{Parameters:} 
 Forget factor \texttt{ff}\;
 \textbf{Initialization:} \\
 Initial data matrix $\mathbf{A}_0 \in \mathbb{R}^{M \times B}$, \\ 
 where $B$ is the number of snapshots per batch\;
 I1. Perform QR-decomposition : $\mathbf{A}_{0} = \mathbf{Q}\mathbf{R}$ \;
 I2. Perform SVD of $\mathbf{R} = \mathbf{U}' \mathbf{D}_0 \mathbf{V}_{0}^{\text{T}}$ and obtain $\mathbf{U}_{0} = \mathbf{Q} \mathbf{U}'$ \;
 \While{New data $\mathbf{A}_i$ available}{
  1. Compute QR decomposition after column-wise concatenation of new data: $\left[\texttt{ff} \cdot \mathbf{U}_{i-1} D_{i-1} \mid \mathbf{A}_{i}\right] = \mathbf{U}_{i-1}^{\prime} \mathbf{D}_{i-1}^{\prime}$\;
  2. Compute SVD of $\mathbf{D}_{i-1}' = \tilde{\mathbf{U}}_{i-1} \tilde{\mathbf{D}}_{i-1} \tilde{\mathbf{V}}^{T}$\;
  3. Preserve the first $K$ columns of $\tilde{\mathbf{U}}_{i-1}$ and denote $\hat{\mathbf{U}}_{i-1}$ \;
  4. Obtain the updated left singular vectors: $\mathbf{U}_i = \mathbf{U}_{i-1}' \hat{\mathbf{U}}_{i-1}$ \;
  5. Truncate to retain $K$ values of $\tilde{\mathbf{D}}_{i-1}$ to obtain $\mathbf{D}_i$ \;
 }
 \caption{Streaming singular value decomposition \cite{levy1998sequential}.}
 \label{Algo1}
\end{algorithm}

\vspace{-0.5cm}

\subsection{Approximate partitioned method of snapshots}
The second building block consists of distributing the computation. This is achieved via the approximate partitioned method of snapshots (APMOS), that allows computing distributed left singular vectors. Note that the primary difference of a standard implementation of APMOS is that it does not provide for a batch-wise update of the singular vectors. Instead, each batch has its respective basis vector calculation which may be stored to disk. While this algorithm does not possess the ability to construct a set of bases for the entire duration of the simulation, its distributed nature allows for the construction of a global basis even in the presence of a domain decomposition. This parallelized computation of the SVD was introduced in~\cite{wang2016approximate} and we recall its main algorithm below. First, APMOS relies on the local calculation of the left singular vectors for the data matrix on each rank of the simulation. To construct this data matrix, snapshots of the local data may be collected over multiple timesteps. Each row of this matrix corresponds to a particular grid point and each column corresponds to a snapshot of data at one time instant. The first stage of local operations is thus
\begin{align}
    \mathbf{A}^i = \mathbf{U}^i \boldsymbol{\Sigma}^i \mathbf{V}^{*i}
\end{align}
where $i$ refers to the index of the rank ranging from 1 to $N_r$ (the total number of ranks), $\mathbf{U}^i \in \mathbb{R}^{M_i \times N}$, $\boldsymbol{\Sigma}^i \in \mathbb{R}^{N \times N}$, and $\mathbf{V}^i \in \mathbb{R}^{N \times N}$. Here, $M_i$ refers to the number of grid points in rank $i$ of the distributed simulation. Note that instead of an SVD, one may also perform a method of snapshots approach for computing $\mathbf{V}^i$ at each rank provided $M_i >> N$. A column-truncated subset of the right singular vectors, $\tilde{\mathbf{V}}^i$, and the singular values, $\tilde{\boldsymbol{\Sigma}}^i$, may then be sent to one rank to perform the exchange of global information for computing the POD basis vectors. This is obtained by collecting the following matrix at rank 0 using the MPI gather command
\begin{align}
    \mathbf{W} = \left[ \tilde{\mathbf{V}}^1 (\tilde{\boldsymbol{\Sigma}}^1)^{\text{T}}, ..., \tilde{\mathbf{V}}^{N_r} (\tilde{\boldsymbol{\Sigma}}^{N_r})^{\text{T}} \right].
\end{align}
In this study, we utilize a truncation factor $r_1=$50 columns of $\mathbf{V}_i$ and $\boldsymbol{\Sigma}_{i}$ for broadcasting. Subsequently a singular value decomposition of $\mathbf{W}$ is performed to obtain
\begin{align}
    \mathbf{W} = \mathbf{X} \boldsymbol{\Lambda} \mathbf{Y}^*. 
\end{align}
Given another threshold factor $r_2$ corresponding to the number of columns retained for $\mathbf{X}$, a reduced matrix $\tilde{\mathbf{X}}$ and reduced singular values $\tilde{\boldsymbol{\Lambda}}$ is broadcast to all ranks. The distributed \emph{global} left singular vectors may then be assembled at each rank as follows for each basis vector $j$
\begin{align}
    \tilde{\mathbf{U}}_j^i = \frac{1}{\tilde{\boldsymbol{\Lambda}}_j} \mathbf{A}^i \tilde{\mathbf{X}}_j
\end{align}
where $\tilde{\mathbf{U}}_j^i$ is the $j^{\text{th}}$ singular vector in the $i^{\text{th}}$ rank, $\tilde{\boldsymbol{\Lambda}}_j$ is the $j^{\text{th}}$ singular value and $\tilde{X}_j$ is the $j^{\text{th}}$ column of the reduced matrix $\tilde{\mathbf{X}}$. A default value of $r_2=5$ columns is chosen for our threshold factor for this last stage. We note that the choices for $r_1$ and $r_2$ may be used to balance communication costs and accuracy for this algorithm. Pseudocode~\ref{Algo2} summarizes this procedure, while listing~\ref{LST2} details the Python implementation in PyParSVD. In this study, we utilize APMOS to allow for streaming updates of singular vectors and values within the online-SVD algorithm. We achieve this by implementing a parallelized variant of the QR decomposition \cite{6691583} (step I1 in Algorithm \ref{Algo1}) and the APMOS variant of the SVD (step I2 in Algorithm \ref{Algo1}).
\begin{algorithm}
\SetAlgoLined
\KwResult{The truncated left singular vector matrix $\tilde{\mathbf{U}}^i$ in each rank $i$ of a distributed computation}
 \textbf{Parameters:} \\
 Threshold factors $r_1$ and $r_2$\;
 \textbf{Algorithm:} \\
 Local data matrix $\mathbf{A}^i$ at each rank $i$ of distributed simulation\;
 1. Perform local SVD calculation of local right singular vectors : $\mathbf{A}^i = \mathbf{U}^i \boldsymbol{\Sigma}^i \mathbf{V}^{*i} $. \;
 2. Truncate $\mathbf{V}^i$, $\boldsymbol{\Sigma}^i$ by retaining only $r_1$ columns to obtain $\tilde{\mathbf{V}}^i$ and $\tilde{\boldsymbol{\Sigma}}^i$ respectively. \;
 3. Obtain $\mathbf{W} = \left[ \tilde{\mathbf{V}}^1 (\tilde{\boldsymbol{\Sigma}}^1)^{\text{T}}, ..., \tilde{\mathbf{V}}^{N_r} (\tilde{\boldsymbol{\Sigma}}^{N_r})^{\text{T}} \right]$ at rank 0 using MPI Gather.\;
 4. Perform SVD, $\mathbf{W} = \mathbf{X} \boldsymbol{\Lambda} \mathbf{Y}^*$ at rank 0.\;
 5. Truncate $\mathbf{X}, \boldsymbol{\Lambda}$ by retaining only $r_2$ columns to obtain $\tilde{\mathbf{X}}, \tilde{\boldsymbol{\Lambda}}$ respectively.\;
 6. Send $\tilde{\mathbf{X}}, \tilde{\boldsymbol{\Lambda}}$ to each rank using MPI Broadcast.\;
 7. Obtain local partition of $j^{\text{th}}$ global left singular vector $\tilde{\mathbf{U}}_j^i = \frac{1}{\tilde{\boldsymbol{\Lambda}}_j} \mathbf{A}^i \tilde{\mathbf{X}}_j$ where $j$ corresponds to the column of the respective matrices.
\caption{Distributed singular value decomposition \cite{wang2016approximate}}.
\label{Algo2}
\end{algorithm}
\subsection{Randomized linear algebra}
The final building block is the randomized linear algebra. To this end, we first define the full SVD of $\mathbf{A}$ as $\mathbf{U} \boldsymbol{\Sigma} \mathbf{V}^*.$ However, this SVD is expensive to evaluate thus we desire a low-rank factorization of $\mathbf{A}$. To evaluate the low rank factorization, we compute an approximate basis for the range of the matrix that is to be factorized. set
\begin{equation}
    \begin{aligned}
       \mathbf{A}_r \approx \mathbf{Q} \mathbf{Q}^{*} \mathbf{A},
    \end{aligned}
\end{equation}
where $\mathbf{A}_r$ is the $r-$rank approximation and the matrix $\mathbf{Q}$ facilitates this approximation. Note if $\mathbf{A} \in \mathbb{R}^{N \times M^{(i)}}$, $\mathbf{Q} \in \mathbb{R}^{N \times r}$ with $r << N$, this low-rank factorization of $\mathbf{A}$ may be expressed as
\begin{equation}
    \begin{aligned}
        \mathbf{A}_r  \approx \mathbf{Q} \tilde{\mathbf{A}},
    \end{aligned}
\end{equation}
where $\tilde{A} = \mathbf{Q}^{*}\mathbf{A}$. We may now perform the SVD of the smaller matrix $\tilde{\mathbf{A}}$, given by
\begin{equation}
    \begin{aligned}
        \tilde{\mathbf{A}} = \tilde{\mathbf{U}}_r \boldsymbol{\Sigma}_r \mathbf{V}_r,
    \end{aligned}
\end{equation}
where $\boldsymbol{\Sigma}_r$ and $\mathbf{V}_r$ are the approximate singular values and the right singular vectors. The left singular vectors of $\mathbf{A}_{r}$ can be recovered by setting
\begin{equation}
    \begin{aligned}
        \mathbf{U}_r = \mathbf{Q} \tilde{\mathbf{U}}_r.
    \end{aligned}
\end{equation}
Finally, we will write the $r-$rank SVD of $\mathbf{A}$ as
\begin{equation}
    \begin{aligned}
        \mathbf{A}_r = \mathbf{U}_r \boldsymbol{\Sigma}_r \mathbf{V}_r
    \end{aligned}
    \label{eq:low_r}
\end{equation}
$\mathbf{Q}$ is generally randomly sampled from a zero-mean unit-variance Gaussian distribution every time a randomized SVD is required. In our implementation, any SVD requirement, for instance in local or global computations may be randomized using the aforementioned process.

\section{Implementation in Python}
\label{S:4}
The implementation in Python has been performed using a factory design pattern. In particular, we define a base class, namely \texttt{Parsvd\_Base} that implements functions shared across the two derived classes \texttt{Parsvd\_Serial} and \texttt{Parsvd\_Parallel}. We also provide a convenient post-processing module that implements the visualization of singular values and SVD modes, namely \texttt{postprocessing}. The routines within this module are linked with the base class \texttt{Parsvd\_Base}. Hence, they can be called directly from the class object, if required. Yet, the functions implemented in the \texttt{postprocessing} module can be also called separately. In the next two subsections, we highlight the key implementation blocks for both the serial and parallel algorithms that are part of the library PyParSVD.

\subsection{Serial implementation}
The two steps of the serial implementation framework consist of data initialization and streaming SVD computation. These are depicted in Listing~\ref{LST1}. The \texttt{initialize} function initializes the problem with the first data segment, while the \texttt{incorporate\_data} function ingests new data in a streaming manner. The initialization step performs a serial SVD, while the streaming step concatenates the data (using the forget factor \texttt{ff}).
\begin{python}[caption={Initialization and streaming steps for the serial algorithm.}, label=LST1]
	def initialize(self, A):
		q, r = np.linalg.qr(A)
        ui, self._singular_values, self.vit = np.linalg.svd(r)
		self._modes = np.matmul(q, ui)[:,:self._K] 
		self._singular_values = \
		    self._singular_values[:self._K]
		return self

	def incorporate_data(self, A):
		m_ap = self._ff * np.matmul(self._modes, \ 
		    np.diag(self._singular_values))
		m_ap = np.concatenate((m_ap, A), axis=-1)
		udashi, ddashi = np.linalg.qr(m_ap)
		utildei, dtildei, vtildeti = np.linalg.svd(ddashi)
		max_idx = np.argsort(dtildei)[::-1][:self._K]
		self._singular_values = dtildei[max_idx]
		utildei = utildei[:,max_idx]
		self._modes = np.matmul(udashi, utildei)
		return self
\end{python}

\subsection{Parallel implementation}
The two key steps of the parallel implementation consists of data initialization and streaming, similarly to the serial algorithm, and are depicted in Listing~\ref{LST2}. We note that the \texttt{initialize} function initializes the problem with the first data segment, while the \texttt{incorporate\_data} function ingests new data in a streaming manner. The initialization step performs a parallel SVD, while the streaming step concatenates the data (using the forget factor \texttt{ff}), and performs a parallel QR decomposition, as described in section~\ref{S:2}. 
\begin{python}[caption={Streaming randomized parallel SVD top-level routines.}, label=LST2]
    def initialize(self, A):
        self.ulocal, self._singular_values = self.parallel_svd(A)
        self._gather_modes()
        return self

    def incorporate_data(self, A):
        self._iteration += 1
        ll = self._ff * 
             np.matmul(self.ulocal, np.diag(self._singular_values))
        ll = np.concatenate((ll, A), axis=-1)
        qlocal, utemp, self._singular_values = self.parallel_qr(ll)
        self.ulocal = np.matmul(qlocal, utemp)
        self._gather_modes()
        return self
\end{python}

The parallel SVD is depicted in Listing~\ref{LST3}, where we perform an SVD (or a low-rank SVD, if required), and we broadcast the solution, before using the APMOS method at each local MPI rank. 
\begin{python}[gobble=2, caption={Parallel SVD.}, label=LST3]
    def parallel_svd(self, A):
        vlocal, slocal = generate_right_vectors(A,self._K)

        # Find Wr
        wlocal = np.matmul(vlocal, np.diag(slocal).T)

        # Gather data at rank 0:
        wglobal = self.comm.gather(wlocal, root=0)

        # perform SVD at rank 0:
        if self.rank == 0:
            temp = wglobal[0]
            for i in range(self.nprocs-1):
                temp = np.concatenate((temp, wglobal[i+1]), axis=-1)
            wglobal = temp

            if self._low_rank:
                x, s = low_rank_svd(wglobal, self._K)
            else:
                x, s, y = np.linalg.svd(wglobal)
        else:
            x = None
            s = None
        x = self.comm.bcast(x, root=0)
        s = self.comm.bcast(s, root=0)

        # perform APMOS at each local rank
        phi_local = []
        for mode in range(self._K):
            phi_temp = 1.0 / s[mode] * \ 
                        np.matmul(A, x[:,mode:mode+1])
            phi_local.append(phi_temp)
        temp = phi_local[0]
        for i in range(self._K - 1):
            temp = np.concatenate((temp, phi_local[i+1]), axis=-1)
        return temp, s[:self._K]
\end{python}

The parallel QR decomposition is depicted in Listing~\ref{LST4}, where we perform an SVD (or a low-rank SVD, if required), and we broadcast the solution, before using the APMOS method at each local MPI rank. 
\begin{python}[gobble=2, caption={Parallel QR decomposition.}, label=LST4]
    def parallel_qr(self, A):

        # Perform local QR
        q, r = np.linalg.qr(A)
        rlocal_shape_0 = r.shape[0]
        rlocal_shape_1 = r.shape[1]

        # Gather data at rank 0:
        r_global = self.comm.gather(r, root=0)

        # perform SVD at rank 0:
        if self.rank == 0:
            temp = r_global[0]
            for i in range(self.nprocs-1):
                temp = np.concatenate(
                    (temp, r_global[i+1]), axis=0)
            r_global = temp
            qglobal, rfinal = np.linalg.qr(r_global)
            qglobal = -qglobal # Trick for consistency
            rfinal = -rfinal

            # For this rank
            qlocal = np.matmul(q, qglobal[:rlocal_shape_0])

            # send to other ranks
            for rank in range(1, self.nprocs):
                self.comm.send(qglobal[rank*rlocal_shape_0:\
                                (rank+1)*rlocal_shape_0],
                               dest=rank, tag=rank+10)

            # Step b of Levy-Lindenbaum - small operation
            if self._low_rank:
                # Low rank SVD
                unew, snew = low_rank_svd(rfinal, self._K)
            else:
                unew, snew, _ = np.linalg.svd(rfinal)
        else:
            # Receive qglobal slices from other ranks
            qglobal = self.comm.recv(source=0, tag=self.rank+10)

            # For this rank
            qlocal = np.matmul(q, qglobal)

            # To receive new singular vectors
            unew = None
            snew = None
        unew = self.comm.bcast(unew, root=0)
        snew = self.comm.bcast(snew, root=0)
        return qlocal, unew, snew
\end{python}

\subsection{Experiments}
\begin{figure*}[b]
    \centering
    \subfigure[Mode 1]{\includegraphics[width=0.32\textwidth]{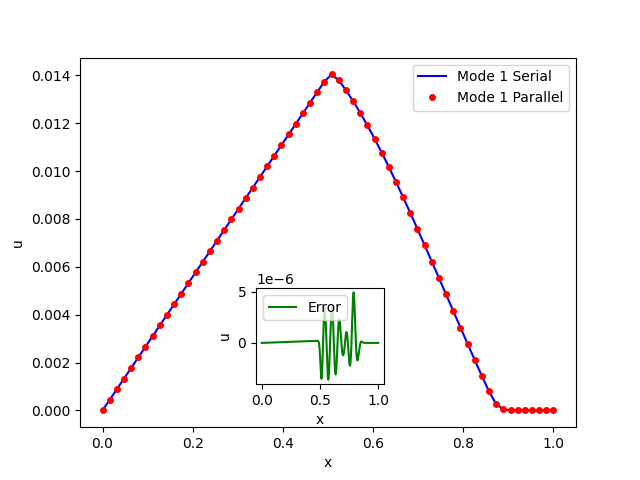}\label{fig:Burgers_1}}
    \subfigure[Mode 2]{\includegraphics[width=0.32\textwidth]{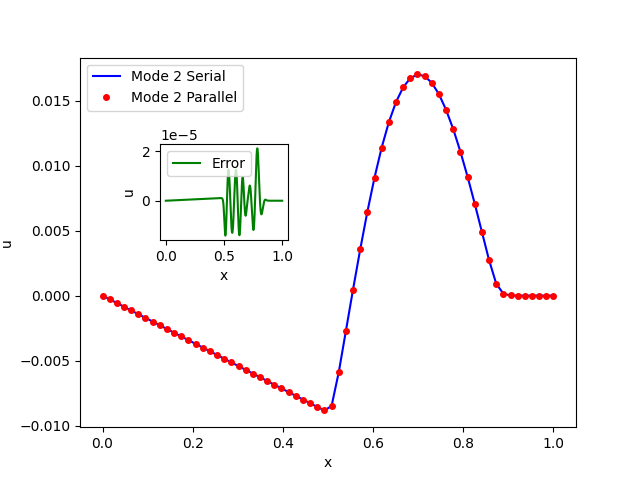}\label{fig:Burgers_2}}
    \subfigure[Weak scaling on Theta]{\includegraphics[width=0.32\textwidth]{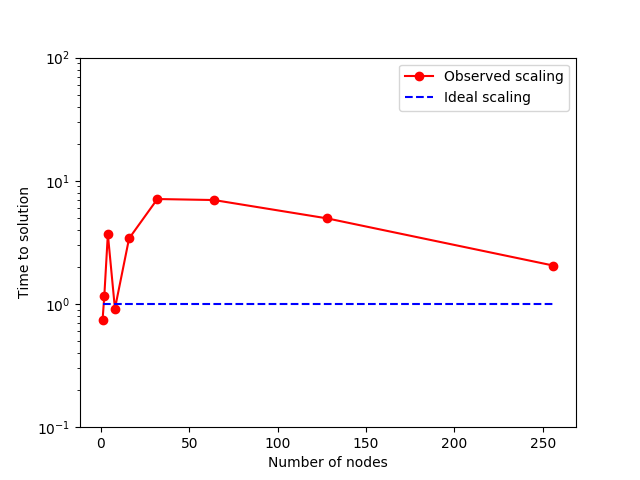}\label{fig:weak_scaling}}
    
    \caption{(a),(b) - Comparison of serial and randomized+parallel deployments of the singular value decomposition for computing coherent structures in the evolution of a viscous Burgers equation. (c) - Weak scaling upto 256 nodes on Theta.}
\end{figure*}

In this section, we outline some experiments to test the accuracy of our proposed algorithm for canonical modal decomposition test cases. Our first test case is given by the viscous Burgers equation, i.e., 
\begin{align}
    \frac{\partial u}{\partial t} + u \frac{\partial u}{\partial x} = \nu \frac{\partial^2 u}{\partial x^2}
\end{align}
with initial conditions given by $u(x,0) = \frac{x}{1 + \sqrt{\frac{1}{t_0}} \exp{(Re \frac{x^2}{4})}}$ and boundary conditions given by $u(0,t)= 0, u(L,t) = 0$. The system is defined on a domain $x \in [0,L] \subset \mathbb{R}^1$ and $t \in [0,t_{f}] \subset \mathbb{R}^1$. The domain length $L$ is set to 1 and the final time of the system is given by 2. A grid-resolution of 16384 points is used for all the assessments on this system. Here, $t_0 = \exp(Re/8)$, and $Re=1/\nu$ is the Reynolds number for the system kept fixed at 1000. This system possesses an analytical solution given by
\begin{align}
    u(x,t) = \frac{\frac{x}{t+1}}{1 + \sqrt{\frac{t+1}{t_0}} \exp{Re \frac{x^2}{4t+4}}  }
\end{align}
and is directly used to generate snapshots for constructing our data matrix. For our first experiment we generate 800 snapshots of data and are therefore tasked with performing the SVD of a data matrix spanning 800 columns and 16384 rows. Figures \ref{fig:Burgers_1} and \ref{fig:Burgers_2} show the validation for a parallel singular vector computation using 4 ranks compared with a serial evaluation for the first and second singular vectors respectively. It is observed that accurate results are obtained with a low error magnitude demonstrating the validity of our algorithm. Preliminary weak scaling results are shown in Figure \ref{fig:weak_scaling} for the same problem explained above with 1024 grid-points per rank of Theta, a leadership class computing resource with Intel Knights-Landing nodes. We assess scaling for upto 256 nodes of Theta, where it is observed that, as the numbers of ranks increase, scaling is seen to follow the ideal trend appropriately. We note that this experiment solely assessed the parallelized and randomized SVD without the utilization of the streaming operation. 

\begin{figure}
    \vspace{-0.45cm}
    \centering
    \mbox{
    \subfigure[Mode 1]{\includegraphics[trim={3cm 2cm 3cm 1cm},clip,width=0.88\textwidth]{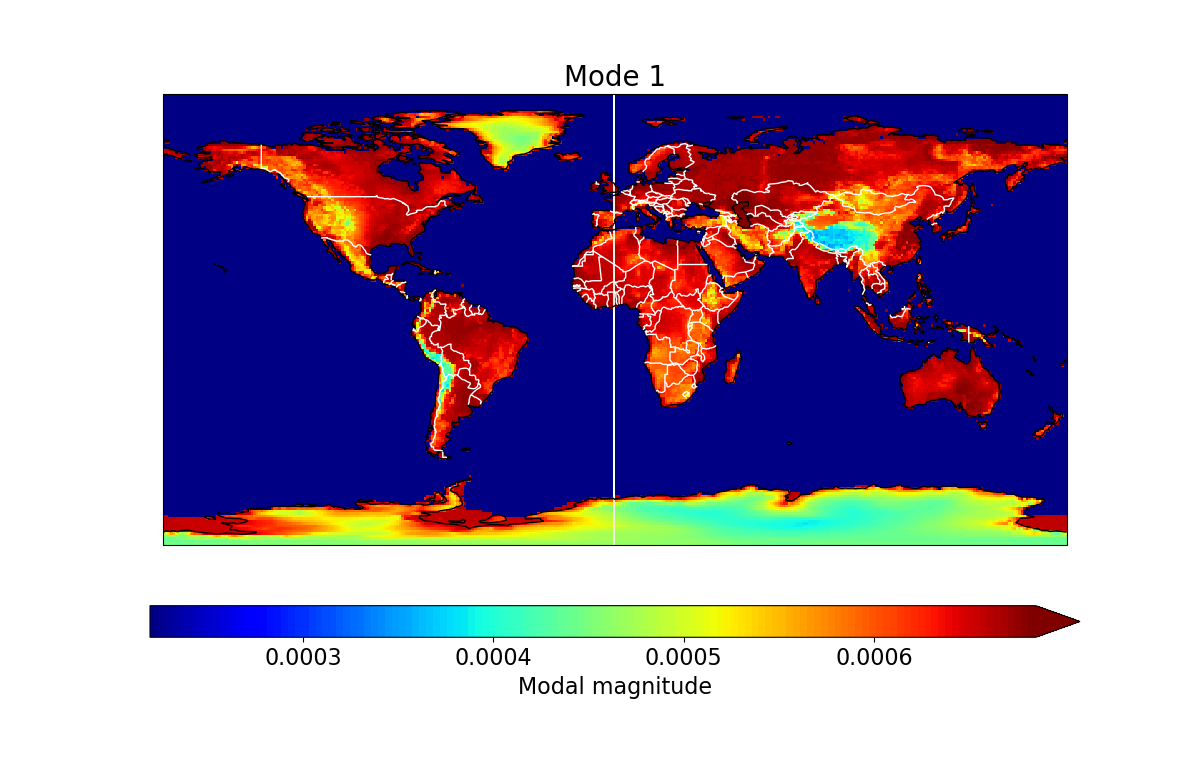}}
    } \\
    \mbox{
    \subfigure[Mode 2]{\includegraphics[trim={3cm 2cm 3cm 1cm},clip,width=0.88\textwidth]{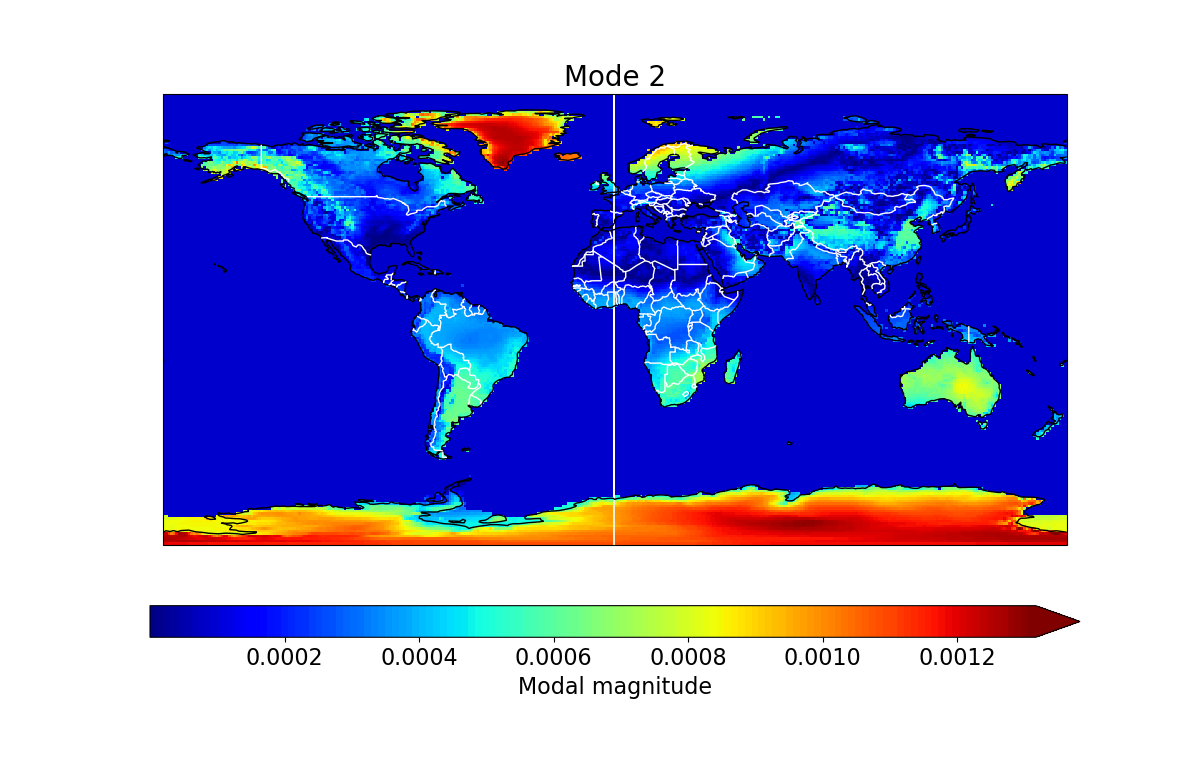}}
    }
    \caption{Modes for the ECMWF surface pressure data set (ERA5) obtained between for the time-period: January 1, 2013 to December 31 2020, at midnight.}
    \label{fig:era_5_modes}
\end{figure}

Finally, as a science application, we demonstrate the utilization of our library, with parallel-IO using NetCDF4, to compute the coherent structures in the ERA5 pressure data set for snapshots obtained from a time-period of January 1, 2013 to December 31 2020 at 6-hourly intervals. Extracting coherent structures in the global pressure is of key importance for predictive modeling and analyses in the context of weather and climate research. Here we demonstrate how the proposed library may effectively extract scientific insight for domain specialists at scale on large compute resources. Coherent structure extraction is displayed for the first two modes in Figure \ref{fig:era_5_modes} .

\section{Conclusions \& Future work}
\label{S:5}

In this study, we have introduced a new Python package for parallelized SVDs of large data matrices. Our algorithm extends the approximate partitioned method of snapshots method for distributed SVD computation with the use of the randomized algorithms and online updates of singular vectors and values for streaming data. Our library is parallelized using MPI and randomized variants of the SVD and QR algorithms are used to accelerate linear algebra. We test our framework for coherent structure extraction on canonical problems for a nonlinear dynamical system, given by the viscous Burgers equation, and a real-world data set given by the global pressure obtained from the ERA5 data set. We validate the fidelity of the proposed package with a serial comparison and also demonstrate weak scaling upto 256 nodes of Theta, a leadership class computing resource with Intel Knights-Landing nodes. Our next steps include the deployment of the devised library on other resources with heterogeneous hardware such as Graphical Processing (GPU) units. Identified steps for accomplishing this include the use of PyCUDA for accelerating linear algebra which may be used to deploy the Randomized SVD and QR algorithms on GPUs. PyCUDA is compatible with MPI4PY and can therefore be used to perform parallelized computations across multiple GPUs.

\footnotesize{
\section*{Acknowledgements}
This material is based upon work supported by the U.S. Department of Energy (DOE), Office of Science, Office of Advanced Scientific Computing Research, under Contract DE-AC02-06CH11357. This research was funded in part and used resources of the Argonne Leadership Computing Facility, which is a DOE Office of Science User Facility supported under Contract DE-AC02-06CH11357. This paper describes objective technical results and analysis. Any subjective views or opinions that might be expressed in the paper do not necessarily represent the views of the U.S. DOE or the United States Government. GM acknowledges support from MOE/NUS Startup grant R-265-000-A36-133. 
}


\bibliographystyle{unsrt}
\bibliography{sample}

\end{document}